# Effects of hydrogen pressure on hydrogenated amorphous silicon thin films prepared by low-temperature reactive pulsed laser deposition

A. Mellos, M. Kandyla[1], D. Palles, and M. Kompitsas

*Theoretical and Physical Chemistry Institute, National Hellenic Research Foundation, 48 Vasileos Constantinou Avenue, 11635 Athens, Greece*

**Abstract**

We deposit intrinsic hydrogenated amorphous silicon (a-Si:H) thin films by reactive pulsed laser deposition, for various hydrogen pressures in the 0 – 20 Pa range, at a low deposition temperature of 120°C, and investigate the hydrogen incorporation, structural, optical, and electrical properties of the films, as a function of the ambient hydrogen pressure. The film thickness decreases linearly as the hydrogen pressure increases. The hydrogen content of the films is determined by infrared spectroscopy and the optical bandgap from UV-Vis-NIR transmittance and reflectance measurements. Electric measurements yield the dark conductivity of the films. The hydrogen concentration of the films lies in the $10^{21} - 10^{22}$ cm$^{-3}$ range and increases with the hydrogen pressure until the latter reaches 15 Pa, beyond which the hydrogen concentration decreases. The optical bandgap and dark conductivity follow the hydrogen concentration variation of the films. The dark conductivity lies in the $10^{-9} - 10^{-10}$ S/cm range. An unusually wide optical bandgap of 2.2 – 2.6 eV is observed.

**Keywords:** hydrogenated amorphous silicon, pulsed laser deposition, wide optical bandgap, conductivity, infrared spectroscopy

---

[1] kandyla@eie.gr

# 1. Introduction

Hydrogenated amorphous silicon (a-Si:H) thin-films are being developed by several methods, including plasma-enhanced chemical vapor deposition [1,2,3], hot-wire chemical vapor deposition [4,5], sputtering [6,7,8], chemical annealing [9], and reactive pulsed laser deposition (PLD) [10,11,12,13], among others, resulting in various optical and electrical properties, film qualities and uniformities, and deposition rates. Reactive pulsed laser deposition allows for the development of hydrogenated silicon thin films via laser ablation of a silicon target in a hydrogen atmosphere. It does not require the use of silane gas, thus reducing the fabrication cost significantly. PLD is a versatile technique because one or more of the deposition parameters (laser fluence, gas pressure, substrate temperature, *etc.*) can be easily adjusted in order to effectively control the film properties. Because it relies on the generation of atomic species with high kinetic energy, due to the high photon energy laser sources typically employed, it allows for low substrate temperatures during fabrication, as opposed to most of the alternative deposition methods, which require deposition temperatures exceeding $200^{o}C$ in order to achieve improved film properties [14,15]. Low processing temperatures are often required in microelectronic fabrication and play a decisive role on the compatibility of a physical or chemical process with microelectronic engineering. A limited number of research studies on PLD fabrication of a-Si:H thin films investigate either the hydrogenation mechanism [12] and hydrogen content [10,12,13] or the crystalline structure [10,11] or the electric [12] and optical [13] properties of the resulting films. However, to the best of our knowledge, a single study to investigate all the aforementioned properties combined for the same sample has evaded the literature. Furthermore, the PLD-grown a-Si:H films reported in the literature have been deposited on substrates kept at room temperature, even though heating of the substrate is known to improve the properties of the deposited films [16,17].

In this work, we investigate the hydrogen incorporation, structural, optical, and electrical properties of pulsed-laser deposited intrinsic a-Si:H thin films, as a function of the hydrogen pressure during deposition. The hydrogen pressure varies in the 0 – 20 Pa range, which is at the lower end of the hydrogen pressures employed for PLD growth of a-Si:H films in previous works. We employ a low substrate temperature (120$^o$C), which is compatible with flexible, low-cost plastic substrates. From infrared spectroscopy, we find the hydrogen content of the films to fall within the $10^{21} – 10^{22}$ cm$^{-3}$ range. The hydrogen content increases with the hydrogen pressure until the latter reaches 15 Pa, beyond which the hydrogen content decreases, due to removal of hydrogen atoms from the surface of the depositing film. The optical bandgap of the films follows the hydrogen concentration variation, maximizing for the highest hydrogen concentration. The a-Si:H films show an unusually wide optical bandgap of 2.2 – 2.6 eV. The dark conductivity of the films also follows the variation of the hydrogen concentration and lies in the $10^{-9} – 10^{-10}$ S/cm range.

## 2. Materials and methods

Hydrogenated amorphous silicon thin films were deposited by reactive pulsed laser deposition [18,19]. Intrinsic silicon wafers were employed as targets for laser irradiation in a high vacuum chamber, in the presence of a hydrogen atmosphere. Prior to deposition, the chamber was evacuated to a base pressure of $10^{-5}$ mbar. A laser beam from a Q-switched Nd:YAG laser system (pulse duration 10 ns, wavelength 355 nm) was focused on the silicon targets, resulting in a fluence of 13 J/cm$^2$. The silicon targets were mounted on a motorized XY translation stage and were raster-scanned during irradiation, which lasted for 3 hrs. A hydrogen gas flow of 10, 15, and 20 Pa dynamic pressure was established in the vacuum chamber during deposition. As a result, thin hydrogenated silicon films were deposited on a substrate (glass or silicon), placed across the silicon targets at a distance of 50 mm, heated to 120°C during film growth. For reference, pure thin silicon films, without hydrogen content, were also deposited on glass or silicon substrates in the presence of 5 Pa dynamic pressure of inert argon gas, keeping the other experimental conditions the same. These samples are referred to as being deposited under 0 Pa of hydrogen pressure in the remaining of the text.

The thickness of the films was measured by an Alpha-Step surface profilometer. The structure of the films was investigated by a X-ray diffractometer with Cu $K_a$ radiation, in the $25° - 60°$ range. Infrared (IR) transmittance measurements, referenced to atmospheric air, were obtained on thin hydrogenated silicon films deposited on bulk silicon substrates, with the aid of a Fourier Transform Infrared (FTIR) spectrometer, operating in the mid-infrared spectral range (500 – 5000 cm$^{-1}$). Optical transmittance and reflectance spectra were recorded with a UV-Vis-NIR spectrophotometer in the wavelength range from 300 to 1200 nm on films deposited on glass substrates. For electrical measurements, hydrogenated silicon films were deposited by pulsed laser deposition on glass substrates covered with a 120-nm gold layer, serving as the back electrode. Subsequently, the chamber was evacuated to the base pressure of $10^{-5}$ Pa to remove hydrogen residues and, without breaking the vacuum, a top 120-nm gold electrode was deposited on the silicon films via pulsed laser deposition. This way there was no silicon oxide layer between the hydrogenated silicon films and both electrodes. Dark conductivity measurements were performed via a two-point probe method, in the geometry shown at the inset of Fig. 6. All measurements were performed at room temperature.

## 3. Results and Discussion

Figure 1 shows the thickness of the a-Si:H films, for various hydrogen pressures in the vacuum chamber during pulsed laser deposition. We observe the film thickness ranges from 160 nm to 560 nm and decreases almost linearly with the hydrogen pressure. X-ray diffraction measurements on the a-Si:H films deposited on glass substrates (not shown here) indicate the films are amorphous. The amorphous structure of the films is in agreement with the strong dependence of the film thickness on the hydrogen pressure and they are both attributed to scattering events between the ejected silicon atoms, which travel from the target to the substrate, and hydrogen. As the hydrogen pressure increases, scattering between silicon atoms and hydrogen becomes more pronounced and results in a smaller amount of less energetic silicon atoms reaching the substrate. The smaller amount of silicon atoms on the substrate results in the formation of a thinner film while the decreased energy of the silicon atoms does not allow for surface displacements that would lead to crystalline film growth.

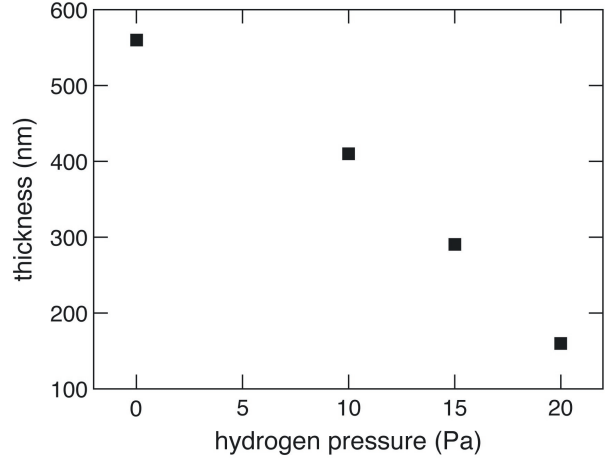

**Figure 1:** Thickness of a-Si:H films as a function of the dynamic hydrogen pressure during pulsed laser deposition.

In order to calculate the hydrogen content of the films, we employ infrared spectroscopy. Figure 2a shows IR transmittance spectra of the a-Si:H films, deposited on 0.35-mm thick silicon substrates, for various hydrogen pressures in the vacuum chamber during pulsed laser deposition. The IR transmittance spectrum of the silicon substrate is shown as well (black curve). The curves are vertically shifted for clarity. The absorption peaks of the silicon substrate spectrum, which are common to all samples, correspond to: (1) a Si-Si vibration mode in bulk c-Si at 611 cm$^{-1}$ [20] (2) a Si-Si vibration mode in the presence of oxygen vacancy in $SiO_2$ at 669 cm$^{-1}$ [21] (3) a two-phonon (LO + LA) Si mode at 738 cm$^{-1}$ [22] (4) a bending Si-O-Si mode at 816 cm$^{-1}$ [21], and (5), (6) a Si-O-Si asymmetric stretching doublet at 1074 and 1200 cm$^{-1}$, with the last two Si-O-Si modes involving mainly oxygen displacements [23]. The absorption peaks at (7) 2341 cm$^{-1}$ and (8) 2360 cm$^{-1}$ correspond to the P and R branch, respectively, of the asymmetric stretching mode of $CO_2$ gas, present in the atmosphere when taking the IR transmittance measurements [24]. Since the reference IR spectrum was acquired in air, the variation of the $CO_2$ doublet intensity in the sample spectra simply depicts the $CO_2$ concentration variation in the sample compartment over time.

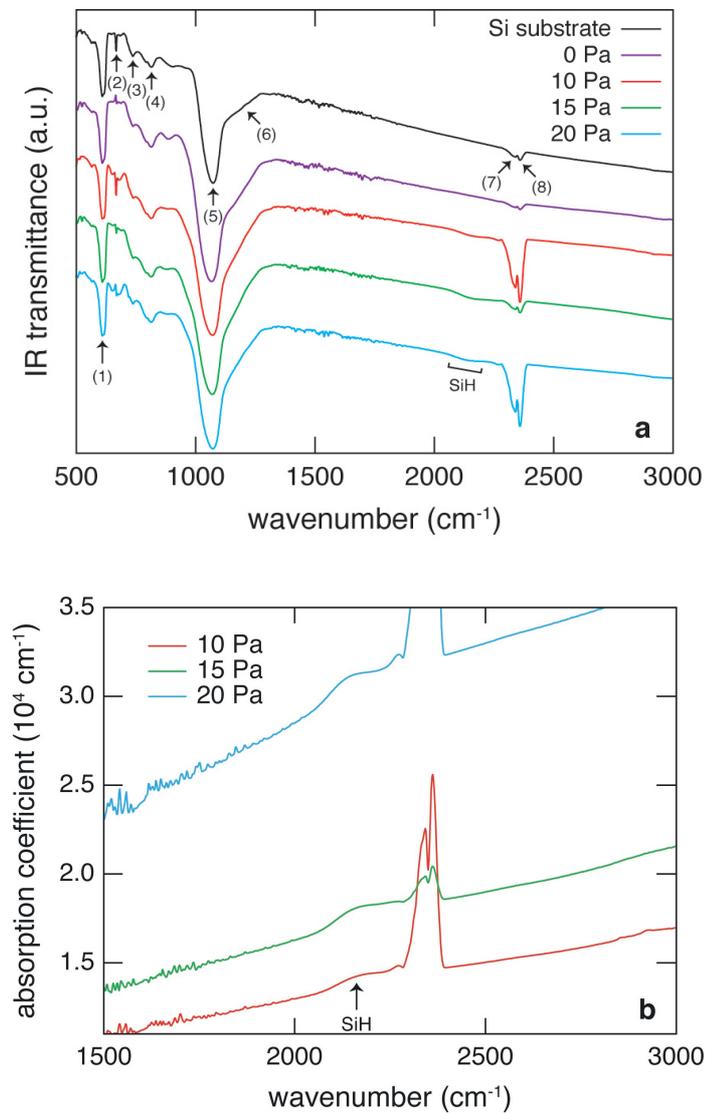

**Figure 2:** (a) Infrared transmittance spectra of a-Si:H films and silicon substrate, for various dynamic hydrogen pressures during pulsed laser deposition. The curves are vertically shifted for clarity. (b) Infrared absorption coefficient in the SiH stretching mode spectral region.

In addition to the absorption peaks present in the spectrum of the silicon substrate, the IR transmittance spectra of the Si:H thin films also show a broad band around ~2100 cm$^{-1}$. The thin silicon film, deposited in argon atmosphere in the absence of hydrogen gas in the vacuum chamber, does not show this band.

Hydrogenated amorphous silicon has a characteristic stretching mode, attributed to clustered monohydrides (SiH) as well as polyhydrides (SiH$_x$, x = 2 or 3), which results in an absorption band centered at 2100 cm$^{-1}$. The hydrogen concentration, $N_H$, can be calculated from the integrated absorbance over this absorption band, $I$, according to the equation:

$$N_H = A_{2100}\, I \quad (1),$$

where

$$I = \int \frac{\alpha_{IR}}{\omega} d\omega \quad (2),$$

$\alpha_{IR}$ is the infrared absorption coefficient, $\omega$ is the frequency in cm$^{-1}$, and $A_{2100} = (2.2 \pm 0.2) \times 10^{20}$ cm$^{-2}$ is a proportionality constant [25]. Figure 2b shows the infrared absorption coefficient for the hydrogenated silicon films in the stretching mode spectral region. In order to extract the absorption coefficient values from IR transmittance measurements, we employ the method described in Ref. [26]. The resulting hydrogen concentration of the films is shown in Fig. 3, where we can see it decreases beyond 15 Pa of hydrogen pressure during deposition, because excessive hydrogen atoms in the atmosphere remove hydrogen atoms from the surface of the depositing film [8].

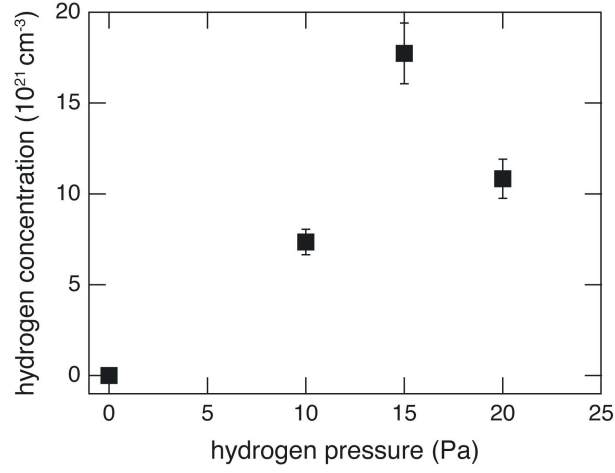

**Figure 3:** Hydrogen concentration of a-Si:H films as a function of the dynamic hydrogen pressure during pulsed laser deposition.

Optical UV-Vis-NIR transmittance, $T$, and reflectance, $R$, measurements (see Supporting Information) allow for the calculation of the bandgap of the Si:H films. From the measured transmittance and reflectance, we calculate the absorption coefficient of the films, $\alpha$, according to the equation:

$$a = \frac{1}{d}\ln\left(\frac{T_g(1-R)}{T}\right) \quad (3),$$

where $d$ is the film thikness and $T_g$ the transmittance of the glass substrate [2] (Fig. 4).

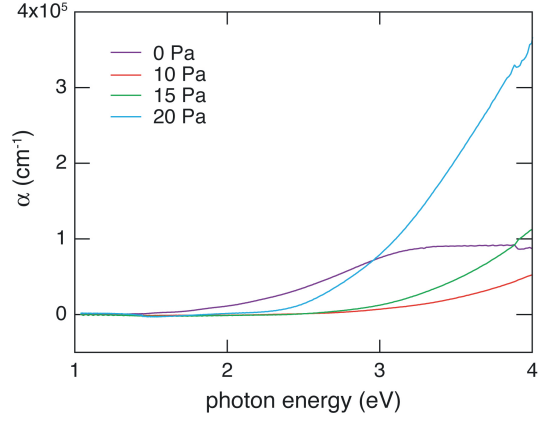

**Figure 4:** Absorption coefficient, *α*, of a-Si:H films for various dynamic hydrogen pressures during pulsed laser deposition.

From a Tauc plot, which depicts $(\alpha h\nu)^{1/2}$ as a function of the photon energy, $h\nu$, we can calculate the optical bandgap of each film as the energy value at which the tangent to the linear part of $(\alpha h\nu)^{1/2}$ inteceptts the x-axis [2,9,28] (see Supporting Information). Figure 5 shows the optical bandgap of the Si:H films, as a function of the hydrogen pressure in the PLD chamber during deposition. We observe the bandgap increases for the hydrogenated films, compared to the pure Si film (0 Pa). Additionally, comparing Fig. 5 with Fig. 3, we note the bandgap increases with increasing hydrogen concentration and then decreases when the hydrogen concentration falls below the maximum value. The presence of hydrogen in Si:H films is known to decrease defect and band-tail states, thus increasing the optical bandgap [27]. The bandgap values shown in Fig. 5 are higher than typical bandgap values for a-Si:H films, which are 1.7 – 1.8 eV [28]. Indeed, depending on the deposition conditions, the bandgap of Si:H films has been observed to deviate from typical values, due to changes in the hydrogen microstructure of the films and the way hydrogen is incorporated in the amorphous matrix [8,28]. Additional reasons for bandgap widening in a-Si:H films include the formation of microvoids [4] and oxygen incorporation [13]. Wide-bandgap a-Si:H films can be used as absorbers with a large open-circuit voltage in multijunction solar cells.

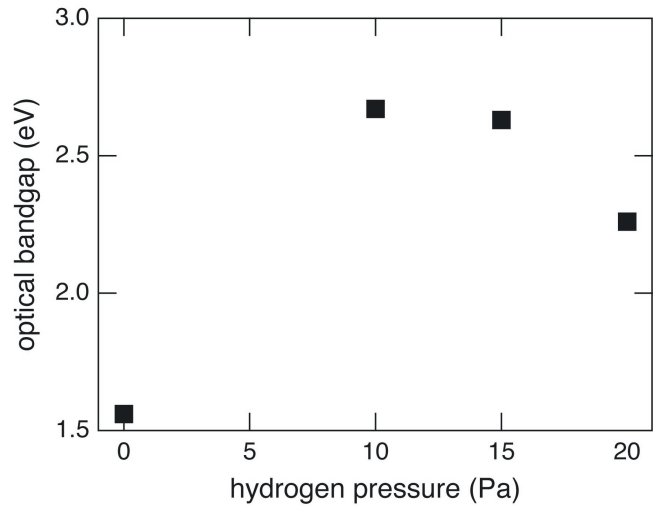

**Figure 5:** Optical bandgap of a-Si:H films as a function of the dynamic hydrogen pressure during pulsed laser deposition.

The dark conductivity of the a-Si:H films is shown in Fig. 6. We observe the incorporation of hydrogen atoms increases the conductivity of the samples. Beyond the hydrogen pressure of 15 Pa during deposition, when the hydrogen concentration of the films decreases, the conductivity decreases as well. Hydrogen passivation of amorphous silicon dangling bonds is known to increase the conductivity of a-Si:H films [27]. Therefore, the conductivity measurements agree with the hydrogen concentration and optical bandgap measurements.

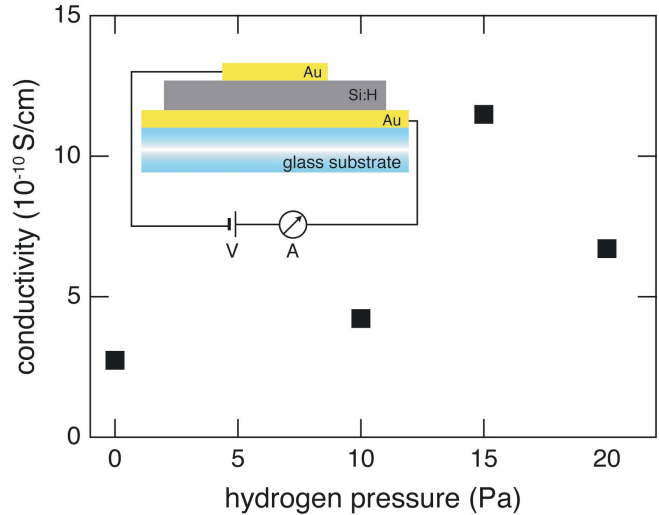

**Figure 6:** Dark electric conductivity of a-Si:H films as a function of the dynamic hydrogen pressure during pulsed laser deposition. Inset: Circuit for electric conductivity measurements.

## 4. Conclusions

We deposited intrinsic a-Si:H thin films by reactive pulsed laser deposition, for various hydrogen pressures in the 0 – 20 Pa range, at a low deposition temperature of 120°C, suitable for microelectronic fabrication on various substrates. The film thickness decreases linearly as the hydrogen pressure increases, due to scattering events between ejected silicon atoms and hydrogen, which results in a smaller amount of silicon atoms reaching the substrate for higher hydrogen pressures. For the formation of crystalline films, we need to employ lower hydrogen pressures, which will allow for more energetic silicon atoms at the substrate. The hydrogen content of the films was determined via infrared spectroscopy and the optical bandgap from UV-Vis-NIR transmittance and reflectance measurements. Electric measurements yield the dark conductivity of the films. The hydrogen concentration decreases beyond 15 Pa of hydrogen pressure during deposition, because excessive hydrogen atoms in the atmosphere remove hydrogen atoms from the surface of the depositing film. The optical bandgap and dark conductivity follow the hydrogen concentration variation of the films because the presence of hydrogen decreases defect and band-tail states, which reduce the bandgap, and passivates silicon dangling bonds, which decrease the conductivity of a-Si:H films. An unusually wide optical bandgap of 2.2 – 2.6 eV is observed.

**Acknowledgements**

Financial support of this work by the General Secretariat for Research and Technology, Greece, under the MS/AC S&T ERA.Net RUS program, STProject-212, FilmSolar, is gratefully acknowledged.

**References**

1. J. Kim, A. J. Hong, J.-W. Nah, B. Shin, F. M. Ross, and D. K. Sadana, ACS Nano **6**, 265 (2012).

2. I. Crupi, S. Mirabella, D. D'Angelo, S. Gibilisco, A. Grasso, S. Di Marko, F. Simone, and A. Terrasi, J. Appl. Phys. **107**, 043503 (2010).

3. F. T. Si, D. Y. Kim, R. Santbergen, H. Tan, R. A. C. M. M. van Swaaij, A. H. Smets, O. Isabella, and M. Zeman, Appl. Phys. Lett. **105**, 063902 (2014).

4. S. R. Jadkar, J. V. Sali, A. M. Funde, N. A. Bakr, P. B. Vidyasagar, R. R. Hawaldar, and D. P. Amalnerkar, Sol. Energ. Mat. Sol. C. **91**, 714 (2007).

5. Y. Kuang, K. H. M. van der Werf, S. Houweling, and R. E. I. Schropp, Appl. Phys. Lett. **98**, 113111 (2011).

6. J. S. Cherng, S. H. Chang, and S. H. Hong, Mater. Res. Bull. **47**, 3036 (2012).

7. D. Girginoudi, C. Tsiarapas, and N. Georgoulas, Appl. Surf. Sci. **257**, 3898 (2011).

8. M. Hossain, H. H. Abu-Safe, H. Naseem, and W. D. Brown, J. Non-Cryst. Solids **352**, 18 (2006).

9. W. Futako, K. Yoshino, C. M. Fortmann, and I. Shimizu, J. Appl. Phys. **85**, 812 (1999).

10. S. Trusso, C. Vasi, and F. Neri, J. Vac. Sci. Technol. A **17**, 921 (1999).

11. H. Fujishiro and S. Furukawa, J. Phys.: Condens. Matter **3**, 7539 (1991).

12. M. Hanabusa and M. Suzuki, Appl. Phys. Lett. **39**, 431 (1981).

13. I. Hanyecz, J. Budai, E. Szilagyi, and Z. Toth, Thin Solid Films **519**, 2855 (2011).

14. F. Meillaud, M. Boccard, G. Bugnon, M. Despeisse, S. Hanni, F.-J. Haug, J. Persoz, J.-W. Schuttauf, M. Stuckelberger, and C. Ballif, Mater. Today **18**, 378 (2015).

15. F.-J. Haug and C. Ballif, Energy Environ. Sci. **8**, 824 (2015).

16. I. Fasaki, M. Kandyla, and M. Kompitsas, Appl. Phys. A **107**, 899 (2012).

17. G. Rijnders and D. H. A. Blank, Growth kinetics during pulsed laser deposition, in: R. Eason (Ed.), Pulsed laser deposition of thin films: applications-led growth of functional materials, (Wiley-Interscience, Hoboken, New Jersey, 2007), chap. 8.

18. I. Fasaki, A. Giannoudakos, M. Stamataki, M. Kompitsas, E. Gyorgy, I. N. Mihailescu, F. Roubani-Kalantzopoulou, A. Lagoyannis, and S. Harissopulos, Appl. Phys. A **91**, 487 (2008).

19. I. Fasaki, M. Kandyla, M. G. Tsoutsouva, and M. Kompitsas, Sensor Actuat. B **176**, 103 (2013).

20. I. V. Belousov, A. Gorchinskiy, P. Lytvyn, G. Kuznetsov, G. Popova, T. Veblaya, D. Zherebeskyy, O. Lysko, O. Vysokolyan, and E. Buzaneva, Mat. Sci. Eng. C **23**, 181 (2003).

21. J. A. Luna-Lopez, G. Garcia-Salgado, T. Diaz-Becerril, J. Carrillo-Lopez, D. E. Vazquez-Valerdi, H. Juarez-Santiesteban, E. Rosendo-Andres, and A. Coyopol, Mat. Sci. Eng. B **174**, 88 (2010).

22. O. Pluchery and J.-M. Costantini, J. Phys. D: Appl. Phys. **45**, 495101 (2012).

23. E. I. Kamitsos, A. P. Patsis, and G. Kordas, Phys. Rev. B **48**, 12499 (1993).

24. A. Oancea, O. Grasset, E. Le Menn, O. Bollengier, L. Bezacier, S. Le Mouelic, and G. Tobie, Icarus **221**, 900 (2012).

25. A. A. Langford, M. L. Fleet, B. P. Nelson, W. A. Lanford, and N. Maley, Phys. Rev. B **45**, 13367 (1992).

26. M. H. Brodsky, M. Cardona, and J. J. Cuomo, Phys. Rev. B **16**, 3556 (1977).

27. N. Savvides, Applications of Surface Science **22/23**, 916 (1985).

28. X. Deng and E. A. Schiff, Amorphous silicon based solar cells, in: A. Luque, S. Hegedus (Eds.), Handbook of Photovoltaic Science and Engineering, John Wiley & Sons, Chichester, 2003, pp. 505 - 65.